# Homophonic Coding Design for Communication Systems Employing the Encoding-Encryption Paradigm

Miodrag J. Mihaljević, Frédérique Oggier and Hideki Imai *

November 13, 2018

**Abstract**

This paper addresses the design of a dedicated homophonic coding for a class of communication systems which, in order to provide both reliability and security, first encode the data before encrypting it, which is referred to as the encoding-encryption paradigm. The considered systems employ error-correction coding for reliability, a stream cipher for encryption, and homophonic coding to enhance the protection of the key used in the stream cipher, on which relies the security of all the system transmissions.

This paper presents a security evaluation of such systems from a computational complexity point of view, which serves as a source for establishing dedicated homophonic code design criteria. The security evaluation shows that the computational complexity of recovering the secret key, given all the information an attacker could gather during passive attacks he can mount, is lower bounded by the complexity of the related LPN (Learning Parity in Noise) problem in both the average and worst case. This gives guidelines to construct a dedicated homophonic encoder which maximizes the complexity of the underlying LPN problem for a given encoding overhead. Finally, this paper proposes a generic homophonic coding strategy that fulfills the proposed design criteria and thus both enhances security while minimizing the induced overhead.

## 1 Introduction

Most communication systems take into account not only the reliability but also the security of the data they transmit. This is particularly true in wireless environments, where the data is inherently more sensible to security threats. Consequently, the system design needs to include both coding schemes for providing error-correction and ciphering algorithms for encryption-decryption. It is common practice to first encrypt the data to ensure the security, and then to encode for reliability purposes. In this paper, we focus on those communication systems which adopt the reverse approach: they first encode the data, and then encrypt it, which we call the *encoding-encryption paradigm*. A famous illustrative example is the most widespread standard for mobile telephony GSM, standing for "Global System for Mobile Communications" (see [4] and [3], for the coding, respectively security details).

In the considered system, encryption-decryption is done through a stream cipher, since the receiver needs to first decrypt the data despite the noise, before performing the decoding, which cannot be done through block ciphers. Consequently the security of the system relies crucially on the private key used in the stream cipher and thus when we refer to the security of systems using the encoding-encryption paradigm, we implicitly mean the security of the keystream generator and the users' secret key.

*Miodrag Mihaljević is with Mathematical Institute, Serbian Academy of Sciences and Arts, Belgrade, Serbia, and with Research Center for Information Security (RCIS), Institute of Advanced Industrial Science and Technology (AIST), Tokyo, Japan. Frédérique Oggier is with Division of Mathematical Sciences, School of Physical and Mathematical Sciences, Nanyang Technological University, Singapore. Hideki Imai is with Faculty of Sciences and Engineering Chuo University, Tokyo, and with Research Center for Information Security (RCIS), Institute of Advanced Industrial Science and Technology (AIST), Tokyo, Japan. Email: miodragm@turing.mi.sanu.ac.rs., frederique@ntu.edu.sg

*Motivation for the Work.* Homophonic coding (see [8] and [9]) is a natural technique to enhance the security of systems employing the encoding-encryption paradigm, since it injects extra randomness in the system, which increases the confusion of a possible adversary by amplifying the channel noise that he experiences. This idea has been recently exploited in [11] and [12], where it was shown from an information-theoretic point of view that with the aid of a dedicated homophonic encoder, the amount of uncertainty that the adversary must face about the secret key given all the information he could gather during different passive or active attacks he can mount, is a decreasing function of the samples available for cryptanalysis. This means that there is a threshold before which the homophonic encoding indeed provides a certain level of unconditional security, but after a large sample is collected, the uncertainty tends to zero, entering a regime in which a computational security analysis is needed for estimating the resistance against the secret key recovery. This paper addresses this computational complexity security evaluation, and highlights how the computational security is related to the homophonic coding design.

*Summary of the Results.* This paper proposes a homophonic code design derived from a security evaluation of the secret key recovery which shows that the security of systems using the encoding-encryption approach can be related to the complexity of solving certain LPN (Learning Parity with Noise) problems. It follows from this analysis that the dedicated homophonic encoding plays a role in securing the system, and that a careful design makes the underlying LPN problem heavily more complex in the average case. This gives guidelines for the design of a dedicated homophonic encoder, that comprises five conditions, two related to the information theoretical security, two regarding the computational security, and one concerning implementation costs. We finally propose a homophonic coding strategy which fulfills all the given criteria.

*Organization of the Paper.* Section 2 introduces the background for the problem addressed in this paper. Section 3 contains the security evaluation from a computational complexity point of view. Implications of the security evaluation on the design of a dedicated homophonic coding are discussed in Section 4, where the code design criteria are established, while the code constructions are given in Section 5. Concluding remarks including some directions for future work are given in Section 6.

## 2 Background

We consider a class of communication systems which, in order to provide both reliability and security, employs the encoding-encryption paradigm, namely: the message is first encoded employing error-correction coding for the purpose of reliability and then encrypted using a stream ciphering. It has been shown in [11] and [12] via an information-theoretic analysis that the use of a dedicated homophonic/wiretap encoding enhances the security of such systems. This section summarizes the reported design and the results of its information-theoretic security evaluation.

### 2.1 System Model

In systems employing the encoding-encryption paradigm (as shown in Figure 1), a stream cipher is used, on which relies the security of the whole system. It is thus crucial to focus on the security of its private key. The system reported and analyzed in [11] and [12] aims at increasing the security of the private key against both passive and active attacks by introducing a dedicated homophonic/wire-tap encoder which involves extra randomness as follows. Let $\mathbf{k}$ be the private key, let $C_H(\cdot)$ denote a homophonic encoder, added at the transmitter, and let
$$\mathbf{u} = [u_i]_{i=1}^{m-\ell} \in \{0,1\}^{m-l}$$

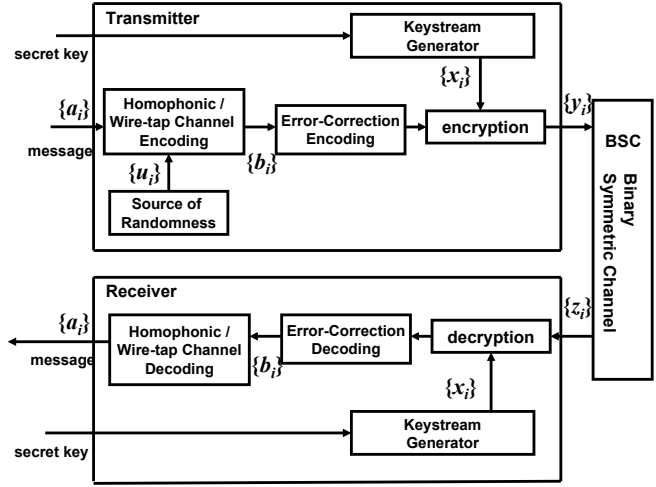

Figure 1: Considered communication system model including homophonic coding.

be a vector of pure randomness[1] where each $u_i$ is the realization of a random variable $U_i$ with distribution $\Pr(U_i = 1) = \Pr(U_i = 0) = 1/2$. The homophonic encoder is positioned before the error-correcting encoder $C_{ECC}(\cdot)$, thus out of the $m$ bits of data to be sent, $m - l$ are replaced by random data, letting actually only $l$ bits

$$\mathbf{a} = [a_i]_{i=1}^{l} \in \{0, 1\}^l$$

of plaintext, to get

$$\mathbf{y} = \mathbf{y}(\mathbf{k}) = C_{ECC}(C_H(\mathbf{a}||\mathbf{u})) \oplus \mathbf{x} \qquad (1)$$

as codeword to be sent, where $\mathbf{x} = \mathbf{x}(\mathbf{k})$ is the output of the keystream generator. We assume that both homophonic and error-correction encoding are linear operations, so that

$$C_H(\mathbf{a}||\mathbf{u}) = [\mathbf{a}||\mathbf{u}]\mathbf{G}_H, \qquad (2)$$

where $\mathbf{G}_H$ is an $m \times m$ matrix, and thus

$$\begin{aligned} C_{ECC}(C_H(\mathbf{a}||\mathbf{u})) &= C_{ECC}([\mathbf{a}||\mathbf{u}]\mathbf{G}_H) \\ &= [\mathbf{a}||\mathbf{u}]\mathbf{G}_H\mathbf{G}_{ECC} \\ &= [\mathbf{a}||\mathbf{u}]\mathbf{G} \end{aligned} \qquad (3)$$

where $\mathbf{G}_{ECC}$ is an $m \times n$ binary generator matrix corresponding to $C_{ECC}(\cdot)$, and $\mathbf{G} = \mathbf{G}_H\mathbf{G}_{ECC}$ is an $m \times n$ binary matrix summarizing the two successive encodings at the transmitter. This homophonic/wire-tap approach is a natural candidate for security enhancement, since homophonic (resp. wire-tap) coding is a well-known technique to create confusion while performing source (resp. channel coding): on the one hand, homophonic coding [9] provides (i) multiple substitutions of a given source vector via randomness so that the coded versions of the source vectors appear as realizations of a random source; (ii) recoverability of the source vector based on the given codeword without knowledge of the randomization. On the other hand, the main goals of wire-tap channel coding (see [14] and [13] when the main channel is error-free) are: (i) amplification of the noise difference between the main and wire-tap channel via randomness; (ii) a reliable transmission in the main channel and at the same time to provide a total confusion of the wire-tapper who observes the communication in the main channel via a noisy channel (wire-tap channel).

---

[1]Although it is assumed for simplicity of presentation, the randomness involved does not need to be "perfect" because the role of randomness is just to enhance security of the employed keystream generator and not to perform encryption itself. In other words, the randomness involved does not need to fulfill all the requirements as the randomness for one-time pad encryption, and accordingly implementation issues become substantially simpler.

In this paper we assume that transmission occurs via a binary symmetric channel (BSC)[2] with known crossover probability $p$, so that the receiver obtains

$$\mathbf{z} = \mathbf{z}(\mathbf{k}) = \mathbf{y} \oplus \mathbf{v} = C_{ECC}(C_H(\mathbf{a}||\mathbf{u})) \oplus \mathbf{x} \oplus \mathbf{v} \tag{4}$$

where $\mathbf{v} = [v_i]_{i=1}^n \in \{0,1\}^n$ and $v_i$ is the realization of a binary random variable $V_i$ such that $\Pr(V_i = 0) = 1 - p$ and $\Pr(V_i = 1) = p$. Since he knows the private key, the receiver starts with the decryption

$$\mathbf{y} = (C_{ECC}(C_H(\mathbf{a}||\mathbf{u})) \oplus \mathbf{x} \oplus \mathbf{v}) \oplus \mathbf{x} = C_{ECC}(C_H(\mathbf{a}||\mathbf{u})) \oplus \mathbf{v}$$

and then decodes

$$C_H(\mathbf{a}||\mathbf{u}).$$

If the decoding is successful, $\mathbf{a}$ is recovered using $C_H^{-1}$ and the receiver informs the transmitter via a feedback link that he could decode. Otherwise he informs the transmitter that retransmission is required.

In the meantime, the adversary keeps on listening and collects over time samples $\{\mathbf{z}^{(t)}\}_{t=1}^\tau$ such that

$$\mathbf{z}^{(t)} = C_{ECC}(C_H(\mathbf{a}^{(t)}||\mathbf{u}^{(t)})) \oplus \mathbf{x}^{(t)} \oplus \mathbf{v}^{(t)} \, , \quad t = 1, 2, ..., \tau \, . \tag{5}$$

employing the notation $\mathbf{a}^{(t)} = [a_i^{(t)}]_{i=1}^\ell$ for the plain text, $\mathbf{u}^{(t)} = [u_i^{(t)}]_{i=1}^{m-\ell}$ for the pure randomness used in the homophonic encoder, $\mathbf{v}^{(t)} = [v_i^{(t)}]_{i=1}^n$ for the channel noise, and $\mathbf{z}^{(t)} = [z_i^{(t)}]_{i=1}^n$ for the received signal.

The main feature of the dedicated homophonic coding used in [11]-[12] is that the encoding is based on randomness and that the legitimate receiver who shares a secret key with the transmitter can perform decoding without knowledge of the randomness employed for the encoding, as shown above. As a result, the decoding complexity without knowledge of the secret key employed in the system approaches the complexity of the exhaustive search for the secret key. The encoding-encryption paradigm for secure and reliable communications enjoys the following desirable properties: (i) when the decryption is performed by bitwise XORing the keystream to the ciphertext, an error in a bit before decryption causes an error in the corresponding bit after decryption, without any error-propagation, and (ii) provides non-availability of the error-free keystream when the communication channel is a noisy one.

## 2.2 Information-Theoretic Security Evaluation

It has been shown in [11] and [12] that from an information-theoretic point of view, it is enough to use a generic wire-tap encoder to increase the security of the system. Namely, under the assumptions that the homophonic encoding matrix $\mathbf{G}_H$ defined in (2) (i) is invertible, so that the receiver can decode the homophonic encoding, and (ii) maps [$\mathbf{a}||\mathbf{u}$] so that in the resulting vector each bit of ciphertext $\mathbf{z}$ is affected by at least one random bit from $\mathbf{u}$, to make sure that each bit of ciphertext is protected, then the homophonic (wire-tap channel) encoding increases the security of the private key $\mathbf{k}$ under both passive and active attacks, where the attacker can modify the ciphertext and learn the effects of these modifications through the feedback link. More precisely

**Theorem 1** *Let*

$$\mathbf{Z}^{(t)} = C_{ECC}(C_H(\mathbf{A}^{(t)}||\mathbf{U}^{(t)})) \oplus f^{(t)}(\mathbf{K}) + \mathbf{V}^{(t)} \, , \quad t = 1, 2, ..., \tau \, , \tag{6}$$

*be the samples available for cryptanalysis obtained over a period of $\tau$ times, where $\mathbf{A}^{(t)} = [A_i^{(t)}]_{i=1}^\ell$, $\mathbf{U}^{(t)} = [U_i^{(t)}]_{i=1}^{m-\ell}$, $\mathbf{V}^{(t)} = [V_i^{(t)}]_{i=1}^n$ and $\mathbf{Z}^{(t)} = [Z_i^{(t)}]_{i=1}^n$, are random variables with entropy $H(A)$, $H(U)$, $H(V)$ and $H(Z)$, representing respectively the data to be sent, pure randomness involved in the wire-tap encoder,*

---

[2]A more general class of binary channels was discussed in [11] and [12].

random noise, and signals seen by the attacker. Let $\mathbf{K}$ be a binary vector random variable uniformly chosen representing the key, such that $H(\mathbf{K}) = |\mathbf{K}|$, and $f^{(j)}(\cdot)$, $j = 1, 2, ..., t$, are given deterministic functions describing the keystream generator, that is, $\mathbf{X}^{(t)} = f^{(t)}(\mathbf{K})$. When $\Pr(V_i^j = 0) \neq \Pr(V_i^j = 1) \neq 1/2$, $i = 1, 2, ..., n$, $j = 1, 2, ..., t$, we have:

$$H(\mathbf{K}|\mathbf{A}, \mathbf{Z}) \begin{cases} > 0 & \text{for} \quad \tau < \tau_{min} \\ \to 0 & \text{for} \quad \tau \geq \tau_{min} \end{cases} \quad (7)$$

for $\mathbf{A} = [\mathbf{A}^{(1)}||\mathbf{A}^{(2)}||...||\mathbf{A}^{(\tau)}]$ and $\mathbf{Z} = [\mathbf{Z}^{(1)}||\mathbf{Z}^{(2)}||...||\mathbf{Z}^{(\tau)}]$.

This theorem statement is interpreted as follows. The uncertainty on the key, measured by the equivocation of the key assuming a known plaintext attack implies, an information-theoretic security assuming that a limited sample only is available for cryptanalysis. This information-theoretic security appears as a consequence of the randomness involved via the homophonic coding as well as the communication channel noise. It has been shown in [11]-[12] that the randomness introduced via homophonic coding has a heavy impact on providing the information-theoretic security. On the other hand, the theorem statement also points out that the information-theoretic security does not hold anymore if an attacker can collect an enough large sample for cryptanalysis, i.e., if the parameter $\tau$ is large enough the equivocation reduces to zero which implies that the secret key is recoverable assuming availability of certain computational power. Accordingly, as long as $\tau < \tau_{thresh}$, the key is protected by the randomness of the homophonic encoder and of the noisy channel in an information-theoretic manner, but that protection cannot last forever if the adversary collects too much data. This does not say that it is easy for the adversary to recover the key, as will be discussed below, but only that the information-theoretic security does not hold anymore.

## 3 Computational Complexity Security Evaluation

This section analyzes the security of the proposed scheme from a computational complexity point of view in the chosen plaintext attacking (CPA) scenario. In this case, the security evaluation consists of establishing how hard it is to find the secret key based on the algebraic representation of the encryption. We will show in our complexity analysis that the hardness of recovering the key relies on the hardness of the LPN problem (see [2],[5], [7], for example). The analysis will pinpoint the features that the homophonic encoder should have so as to create an increased complexity of the underlying LPN problem in the average case.

### 3.1 Preliminaries

We consider the scenario where enough large samples $\{\mathbf{z}^{(t)}\}_{t=1}^{\tau}$ have been recorded by an attacker, who can now try to find the employed secret key $\mathbf{k}$ contained in $\mathbf{x}^t = \mathbf{x}^{(t)}(\mathbf{k})$ using from (5)

$$\mathbf{z}^{(t)} = C_{ECC}(C_H(\mathbf{a}^{(t)}||\mathbf{u}^{(t)})) \oplus \mathbf{x}^{(t)} \oplus \mathbf{v}^{(t)}, \quad t = 1, 2, ..., \tau,$$

since he has a probability of error in recovering the key which now tends to zero.

For the simplicity of exposition, we assume from now on that $|\mathbf{K}| = n$. We further perform the security evaluation under the following two assumptions:

- $\mathbf{x}^{(t)} = f^{(t)}(\mathbf{k}) = \mathbf{k}\mathbf{S}^t$, $t = 1, 2, ..., \tau$, where $\mathbf{S} = [s_{i,j}]_{i=1}^n{}_{j=1}^n$, is a binary matrix, and

$$\mathbf{S}^t = [\mathbf{S}_1^{(t)}, \mathbf{S}_2^{(t)}, ..., \mathbf{S}_n^{(t)}] \quad (8)$$

where each $\mathbf{S}_i^{(t)}$, $i = 1, 2, ..., n$, denotes a column of the $t$th power of the matrix $\mathbf{S}$; note that usually $f^{(t)}(\cdot)$ is a heavily nonlinear function, and its consideration as a linear one actually implies a scenario for evaluation of a lower bound of the complexity for the secret key recovery;

- we consider the chosen plaintext attack where the data is the whole zero vector, i.e. $\mathbf{a}^{(t)} = \mathbf{0}$, for each $t$.

Under the above assumptions, and recalling from (3) that both $C_{ECC}$ and $C_H$ are linear encoders, we can write
$$\mathbf{k}\mathbf{S}^t \oplus [\mathbf{0}||\mathbf{u}^{(t)}]\mathbf{G} = \mathbf{z}^{(t)} \oplus \mathbf{v}^{(t)},$$
from which we obtain an algebraic representation of the recovery problem in terms of a noisy system of linear equations, as seen by the adversary:

$$\begin{bmatrix} \mathbf{k}\mathbf{S}_1^{(t)} \\ \mathbf{k}\mathbf{S}_2^{(t)} \\ \vdots \\ \mathbf{k}\mathbf{S}_n^{(t)} \end{bmatrix} \oplus \begin{bmatrix} [\mathbf{0}||\mathbf{u}^{(t)}]\mathbf{G}_1 \\ [\mathbf{0}||\mathbf{u}^{(t)}]\mathbf{G}_2 \\ \vdots \\ [\mathbf{0}||\mathbf{u}^{(t)}]\mathbf{G}_n \end{bmatrix} = \begin{bmatrix} z_1^{(t)} \\ z_2^{(t)} \\ \vdots \\ z_n^{(t)} \end{bmatrix} \oplus \begin{bmatrix} v_1^{(t)} \\ v_2^{(t)} \\ \vdots \\ v_n^{(t)} \end{bmatrix}, \quad t = 1, 2, ..., \tau, \tag{9}$$

where $\mathbf{u}^{(t)} = [u_i^{(t)}]_{i=1}^{m-\ell}$ and $\mathbf{G}_i$ denotes the $i$th column of $\mathbf{G}$.

**Remark 1** Note that in the set $\{[\mathbf{0}||\mathbf{u}^{(t)}]\mathbf{G}_i\}_{i=1}^n$ all the elements could be split into two non-overlapping subsets such that a subset contains $k$ linearly independent elements, $k$ at most $m - \ell$, and the other subset contains $n - k$ elements each of which is a linear combination of the elements from the first set, since $[\mathbf{0}||\mathbf{u}^{(t)}]\mathbf{G}$ only involves the lower part of $\mathbf{G}$, which is an $(m - l) \times n$ matrix, which has thus at most $m - l$ linearly independent columns, and the other columns can be obtained as linear combinations.

The problem of solving a system of linear equations in the presence of noise is related to the so-called LNP problem, defined formally as follows.

**Definition - The LPN Problem**. Let $\langle \cdot | \cdot \rangle$ denote the binary inner product. Let $\mathbf{s}$ be a random $n$-bit vector, let $\epsilon \in ]0, 1/2[$ be a constant noise parameter, let $\text{Ber}_\epsilon$ be the Bernoulli distribution with parameter $\epsilon$ (so if $\nu \leftarrow \text{Ber}_\epsilon$ then $\Pr(\nu = 0) = \epsilon$ and $\Pr(\nu = 0) = 1 - \epsilon$), and let $A_{\mathbf{s},\epsilon}$ be the distribution defined as

$$\{\mathbf{a} \leftarrow \{0,1\}^n; \nu \leftarrow \text{Ber}_\epsilon : (\mathbf{a}, \langle \mathbf{s} | \mathbf{a} \rangle \oplus \nu)\}$$

Let $\mathcal{A}_{\mathbf{s},\epsilon}$ denote an oracle which outputs independent samples according to the above distribution. An algorithm $\mathcal{M}$ is said to $(q, t, m, \theta)$-solve the $\text{LPN}_{n,\epsilon}$ problem if

$$\Pr[\mathbf{s} \leftarrow \{0,1\}^n : \mathcal{M}^{\mathcal{A}_{\mathbf{s},\epsilon}}(1^n) = \mathbf{s}] \geq \theta ,$$

and furthermore $\mathcal{M}$ runs in time at most $t$, memory at most $m$, and makes at most $q$ queries to its oracle.

What the LPN problem captures is that, given a security parameter $k$, a secret vector $\mathbf{x}$, and $\mathbf{g}_1, ..., \mathbf{g}_n$ randomly chosen binary vectors of length $n = O(k)$, it is possible knowing $y_i = \langle \mathbf{x} | \mathbf{g}_i \rangle$ and $\{\mathbf{g}_i\}_{i=1}^n$ to solve for $\mathbf{x}$ using standard linear-algebraic techniques as long as there is no noise. However, when each $y_i$ is flipped (independently) with probability $p$, finding $\mathbf{x}$ becomes much more difficult. The problem of learning $\mathbf{x}$ in this latter case is refereed to as the learning parity in noise (LPN) problem.

Finally note that the LPN problem is equivalent to the problem of decoding of a general linear block code and it is known that this problem is NP-complete [1], and that relating security of an encryption technique to the LPN problem has been employed for security evaluation of certain stream ciphers (see [10], for example).

## 3.2 Complexity Evaluation

A systematic way to solve a system of linear equations, with or without noise, is to perform a Gaussian elimination. If the system furthermore contains unknowns that we are not interested in finding, it is natural to start by removing them, so as to obtain a system with a smaller number of equations, where only the unknowns we would like to find are left. We will now show how such a strategy changes the noise present in the system of equations.

**Lemma 1** *Consider the following system of $N$ equations over the binary field $GF(2)$ to be solved for $x_1, \ldots, x_L$, $L \leq N$:*

$$(\bigoplus_{j=1}^{L} \alpha_j^{(i)} x_j) \oplus y_i = z_i \oplus e_i \ , \ \ i = 1, 2, ..., M \ ,$$

$$(\bigoplus_{j=1}^{L} \alpha_j^{(i)} x_j) \oplus (\bigoplus_{j=1}^{M} \beta_j^{(i)} y_j) = z_i \oplus e_i \ , \ \ i = M+1, M+2, ..., N \ , \quad (10)$$

*where $\{z_i\}_{i=1}^{N}$, $\{\alpha_j^{(i)}\}_{j=1, i=1}^{L, N}$ and $\{\beta_j^{(i)}\}_{j=1, i=1}^{M, N}$ are known, $\{x_j\}_{j=1}^{L}$, $\{y_j\}_{j=1}^{M}$ and $\{e_i\}_{i=1}^{N}$ are unknown, and each $e_i$ is a realization of a random variable $E_i$ such that $\Pr(E_i = 1) = p < 1/2$, $i = 1, 2, ..., N$. If*

1. *the Hamming weight of each vector $[\beta_1^{(i)}, \ldots, \beta_M^{(i)}]$ is greater or equal to some parameter $w$, for $i = M+1, M+2, ..., N$,*

2. *and no $\bigoplus_{j=1}^{M} \beta_j^{(k)} y_j$, $k \in \{M+1, M+2, \ldots, N\}$, is a linear combination of any other $w$ or less $\bigoplus_{j=1}^{M} \beta_j^{(i)} y_j$, $i \in \{M+1, M+2, \ldots, N\}$, i.e., there are at least $w$ linearly independent sums $\bigoplus_{j=1}^{M} \beta_j^{(i)} y_j$ among those $i \in \{M+1, \ldots, N\}$,*

*then, the problem of recovering the unknown $x_1, x_2, \ldots, x_L$ is the problem of solving the following system of equations:*

$$(\bigoplus_{k=1}^{M} \beta_k^{(i)} (\bigoplus_{j=1}^{L} \alpha_j^{(k)} x_j)) \oplus (\bigoplus_{j=1}^{L} \alpha_j^{(i)} x_j) = z_i \oplus (\bigoplus_{k=1}^{M} \beta_k^{(i)} z_k) \oplus e_i^* \ , \ \ i = M+1, M+2, ..., N \ , \quad (11)$$

*where $e_j^*$ is a realization of a random variable $E_j^*$ such that $\Pr(E_j^* = 1) > p_w = \frac{1-(1-2p)^{w+1}}{2}$.*

*Proof.* For every $i \in \{M+1, M+2, ..., N\}$, adding the following linear combination of the first $M$ equations

$$(\bigoplus_{k=1}^{M} \beta_k^{(i)} (\bigoplus_{j=1}^{L} \alpha_j^{(k)} x_j)) \oplus (\bigoplus_{k=1}^{M} \beta_k^{(i)} y_k) = \bigoplus_{k=1}^{M} \beta_k^{(i)} (z_k \oplus e_k) \ , \quad (12)$$

to the $i$th equations of the system yields:

$$(\bigoplus_{k=1}^{M} \beta_k^{(i)} (\bigoplus_{j=1}^{L} \alpha_j^{(k)} x_j)) \oplus (\bigoplus_{j=1}^{L} \alpha_j^{(i)} x_j) = z_i \oplus (\bigoplus_{k=1}^{M} \beta_k^{(i)} z_k) \oplus e_i \oplus (\bigoplus_{k=1}^{M} \beta_k^{(i)} e_k) \ . \quad (13)$$

We are left to compute the probability $\Pr(E_i^* = 1)$, where

$$E_i^* = E_i \oplus (\bigoplus_{k=1}^{M} \beta_k^{(i)} E_k), \ i = M+1, \ldots, N.$$

Since $i \geq M+1$, $E_i$ is independent of $\beta_k^{(i)} E_k$ for every $1 \leq k \leq M$. We are thus summing the components of the vector

$$[E_i, E_1 \beta_1^{(i)}, \ldots, E_M \beta_M^{(i)}]$$

and

$$\begin{aligned} \Pr(E_i^* = 1) &= 1 - \Pr(E_i^* = 0) \\ &= 1 - \Pr(E_i \oplus (\bigoplus_{k=1}^{M} \beta_k^{(i)} E_k = 0)). \end{aligned}$$

Now the probability that an even number of digits are 1 in a sequence of $M+1$ independent binary digits is [6, Lemma 1]
$$\frac{1+(1-2p)^{M+1}}{2}$$
if $p$ is the probability that every digit is 1. Since
$$\frac{1+(1-2p)^{M}}{2} > \frac{1+(1-2p)^{M+1}}{2}, \ p < 1/2,$$
we have that
$$1 - \frac{1+(1-2p)^{M}}{2} < 1 - \frac{1+(1-2p)^{M+1}}{2},$$
and
$$\begin{aligned}
\Pr(E_i^* = 1) &= 1 - \Pr(E_i \oplus (\bigoplus_{k=1}^{M} \beta_k^{(i)} E_k = 0)) \\
&> 1 - \frac{1+(1-2p)^{w+1}}{2} \\
&= \frac{1-(1-2p)^{w+1}}{2}
\end{aligned}$$

since by assumption 1., the weight of each vector of $[\beta_1^{(i)}, \ldots \beta_M^{(i)}]$ is at least $w$, and according to the assumption 2., there is no linear combination of the equations which can reduce the corruption noise value lower bounded by $p_w$ (i.e., it cannot be reduced via any further linear processing of the system of equations). ■

This leads to the main evaluation result:

**Theorem 2** *The complexity of recovering the secret key* $\mathbf{k}$ *based on the algebraic representation of the scheme is lower bounded by the complexity of solving the $LPN_{n,\epsilon}$ problem where, $\epsilon = \frac{1-(1-2p)^{w+1}}{2}$ and $n$, $w$ and $p$ are the parameters of the scheme, representing resp. the length of the key, a parameter of the homophonic encoder and the probability of the BSC.*

*Proof.* From (9), we have the following system of $\tau n$ overdefined consistent but probabilistic equations over the binary field $GF(2)$:
$$\begin{array}{rcl}
\mathbf{k}\mathbf{S}_1^{(t)} \oplus [\mathbf{0}||\mathbf{u}^{(t)}]\mathbf{G}_1 &=& z_1^{(t)} \oplus v_1^{(t)} \\
\mathbf{k}\mathbf{S}_2^{(t)} \oplus [\mathbf{0}||\mathbf{u}^{(t)}]\mathbf{G}_2 &=& z_2^{(t)} \oplus v_2^{(t)} \\
\vdots & & \\
\mathbf{k}\mathbf{S}_n^{(t)} \oplus [\mathbf{0}||\mathbf{u}^{(t)}]\mathbf{G}_n &=& z_n^{(t)} \oplus v_n^{(t)}
\end{array}, \quad t=1,2,...,\tau \ , \quad (14)$$

where each equation is correct with probability equal to $p$, $\mathbf{0}$ is a $\ell$-dimensional vector of all zeroes, and $\mathbf{u}^{(t)} = [u_i^{(t)}]_{i=1}^{m-\ell}$.

The above system of equations fits the setting of Lemma 1, since we have $N = \tau n$ equations, for $L = n$ unknown, where $\bigoplus_{j=1}^{L} \alpha_j^{(k)} x_j$, $k = 1, \ldots, N$ correspond to $\mathbf{k}\mathbf{S}_i^{(t)}$, $i = 1, \ldots, n$, $t = 1, \ldots, \tau$, and $y_j$, $j = 1, \ldots, M$ together with $\bigoplus_{j=1}^{M} \beta_j^{(k)} y_j$ for $k = M+1, \ldots, N$ correspond to $[\mathbf{0}||\mathbf{u}^{(t)}]\mathbf{G}_i$, $i = 1, \ldots, M$, $t = 1, \ldots, \tau$, since according to Remark 1, we can indeed separate the $\{[\mathbf{0}||\mathbf{u}^{(t)}]\mathbf{G}_i\}_{i=1}^{n}$ for every $t$ into one set of linear independent vectors, and another set which is obtained as linear combinations of the first set ($M$ is then $\tau k$, where $k$ is at most $m-l$).

Note that the above system of $\tau n$ equations contains only $n + \tau(m - \ell)$ unknown variables, and that our goal is to recover $\mathbf{k}$ only, i.e., we do not have any interest in recovering $\{u_i^{(t)}\}_{i=1}^{m-\ell}$, $t = 1, 2, ..., \tau$. Thus, via

Gaussian elimination, we can remove the $\tau(m-l)$ unknown $\{u_i^{(t)}\}_{i=1}^{m-\ell}$, $t = 1, 2, ..., \tau$, and obtain $\tau(n-m+\ell)$ equations where only $\mathbf{k}$ is unknown. This transforms the initial system of $\tau n$ equations into the following one with $\tau(n - m - \ell)$ equations (in total) and $n$ unknowns $\mathbf{k}$:

$$\begin{array}{rcl}
\mathcal{L}_1^{(k)}(\mathbf{k}) & = & \mathcal{L}_1^{(z)}([z_i^{(t)}]_{i=1}^n) \oplus \mathcal{L}_1^{(v)}([v_i^{(t)}]_{i=1}^n) \\
\mathcal{L}_2^{(k)}(\mathbf{k}) & = & \mathcal{L}_2^{(z)}([z_i^{(t)}]_{i=1}^n) \oplus \mathcal{L}_2^{(v)}([v_i^{(t)}]_{i=1}^n) \\
& \vdots & \\
\mathcal{L}_{n-m+\ell}^{(k)}(\mathbf{k}) & = & \mathcal{L}_{n-m+\ell}^{(z)}([z_i^{(t)}]_{i=1}^n) \oplus \mathcal{L}_{n-m+\ell}^{(v)}([v_i^{(t)}]_{i=1}^n)
\end{array}, \quad t = 1, 2, ..., \tau, \quad (15)$$

where $\mathcal{L}_j^{(k)}(\cdot)$, $\mathcal{L}_j^{(z)}(\cdot)$ and $\mathcal{L}_j^{(v)}(\cdot)$, $j = 1, 2, ..., n-m+\ell$, are linear functions, all of them specified by the matrix $\mathbf{G}$ and the Gaussian elimination used to remove the random bits $\mathbf{u}^{(t)}$, while $\mathcal{L}_j^{(k)}(\cdot)$ further depends on the matrix $\mathbf{S}^t$. Note that the Gaussian elimination of the variables $\{u_i^{(t)}\}_{i=1}^{m-\ell}$, can be performed independently for each $t$, implying that the entire complexity (for $t = 1, 2, ..., \tau$) is upper-bonded by $\tau O(n^{2.7})$ assuming employment of the most efficient algorithm for the Gaussian processing.

Lemma 1 and its underlying assumptions provide that each equation in (15) is correct with some probability lower than $1 - p_w$, where
$$p_w = \frac{1 - (1 - 2p)^{w+1}}{2},$$
since the noise $(\mathbf{v}_1^*)^{(t)} = \mathcal{L}_1^{(v)}([v_i^{(t)}]_{i=1}^n), \ldots, (\mathbf{v}_{n-m+\ell}^*)^{(t)} = \mathcal{L}_{n-m+\ell}^{(v)}([v_i^{(t)}]_{i=1}^n)$ has coefficients that are the realization of a random variable which takes value 1 with probability greater than $p_w = \frac{1-(1-2p)^{w+1}}{2}$. The above system of $\tau(n-m+\ell)$ equations can consequently be rewritten as:

$$\begin{array}{rcl}
\mathcal{L}_1^*([k_i]_{i=1}^n) & = & \mathcal{L}_1^{(z)}([z_i^{(1)}]_{i=1}^n) \\
\mathcal{L}_2^*([k_i]_{i=1}^n) & = & \mathcal{L}_2^{(z)}([z_i^{(1)}]_{i=1}^n) \\
& \vdots & \\
\mathcal{L}_{n-m+\ell}^*([k_i]_{i=1}^n) & = & \mathcal{L}_{n-m+\ell}^{(z)}([z_i^{(1)}]_{i=1}^n) \\
\mathcal{L}_{n-m+\ell+1}^*([k_i]_{i=1}^n) & = & \mathcal{L}_1^{(z)}([z_i^{(2)}]_{i=1}^n) \\
\mathcal{L}_{n-m+\ell+2}^*([k_i]_{i=1}^n) & = & \mathcal{L}_2^{(z)}([z_i^{(2)}]_{i=1}^n) \\
& \vdots & \\
\mathcal{L}_{\tau(n-m+\ell)}^*([k_i]_{i=1}^n) & = & \mathcal{L}_{n-m+\ell}^{(z)}([z_i^{(\tau)}]_{i=1}^n)
\end{array} \quad (16)$$

where each equation is incorrect with probability greater than $p_w = \frac{1-(1-2p)^{w+1}}{2}$, and where $\mathcal{L}_j^*$, $j = 1, 2, ..., \tau(n-m+\ell)$, are linear functions.

We finally get:
$$\begin{array}{rcl}
\langle \mathbf{k} | \mathbf{c}_1 \rangle & = & d_1 \\
\langle \mathbf{k} | \mathbf{c}_2 \rangle & = & d_2 \\
& \vdots & \\
\langle \mathbf{k} | \mathbf{c}_{n-m+\ell} \rangle & = & d_{m-n+\ell} \\
\langle \mathbf{k} | \mathbf{c}_{n-m+\ell+1} \rangle & = & d_{m-n+\ell+1} \\
\langle \mathbf{k} | \mathbf{c}_{n-m+\ell+2} \rangle & = & d_{m-n+\ell+2} \\
& \vdots & \\
\langle \mathbf{k} | \mathbf{c}_{\tau(n-m+\ell)} \rangle & = & d_{\tau(m-n+\ell)}
\end{array}, \quad (17)$$

where each equation is correct with a probability upper-bounded by $1 - p_w = 1 - \frac{1-(1-2p)^{w+1}}{2}$, and where the $n$-dimensional binary vectors $\{\mathbf{c}_j\}_{j=1}^{\tau(n-m+\ell)}$ and $\{d_j\}_{j=1}^{\tau(n-m+\ell)}$ are known.

According to the definition of the LPN problem and the above representation, the problem of recovering the secret key is at least as hard as the LPN$_{n,\epsilon}$ problem with $\epsilon = \frac{1-(1-2p)^{w+1}}{2}$, which concludes the proof of the theorem. ∎

## 4 Homophonic Encoder Design Criteria

From the above computational security evaluation, it is clear that the design of the homophonic encoder influences the computational complexity of cryptanalysis. In this section, we explicit code design criteria for homophonic coding, taking into account not only the computational and information theoretical security, but also the implementation complexity. Requirements can be expressed either as a function of $\mathbf{G}_H$ given $\mathbf{G}_{ECC}$, or as a joint function of $\mathbf{G}_H$ and $\mathbf{G}_{ECC}$.

Indeed, the latter holds in the case of a design of the encoding-encryption system from scratch, where the design should include a coding box which performs the concatenation of homophonic and error-correction coding in a manner which fits the rate of the concatenated code to the given constraints. The former on the other hand applies when upgrading existing systems employing the encoding-encryption paradigm, in which case, the implementation assumption is that the employed, already existing, binary linear block error-correction code $(m, n)$ which encodes $m$ bits into a codeword from $GF(2^n)$, could be replaced with a binary block code $(m', n)$ with the same error correction capability but with $m' > m$. Accordingly, $m' - m$ random bits can be concatenated with $m$ information bits and mapped into the new $m'$-bits via a homophonic encoder. The obtained output from the homophonic encoder is the input for the error-correcting one.

We recall first that the basic requirements on the matrix $\mathbf{G}_H$, as far as information theoretical security is concerned, are [11, 12]:

- **Invertibility.** The matrix $\mathbf{G}_H$ should be an invertible matrix, so that the receiver can decode the homophonic encoding.

- **Security.** The matrix $\mathbf{G}_H$ should map [$\mathbf{a}||\mathbf{u}$] so that in the resulting vector each bit of data from $\mathbf{a}$ is affected by at least one random bit from $\mathbf{u}$ (to provide a background that each bit of ciphertext is affected by at least one bit of $\mathbf{u}$).

### 4.1 Computational complexity design criteria

Recall from (3) that

$$\begin{aligned} C_{ECC}(C_H(\mathbf{a}||\mathbf{u})) &= [\mathbf{a}||\mathbf{u}]G_H G_{ECC} \\ &= [\mathbf{a}||\mathbf{u}]\mathbf{G} \end{aligned}$$

where $\mathbf{G} = [g_{i,j}]_{i=1\ j=1}^{m\ \ n}$ is an $m \times n$ matrix containing both the homophonic and the error correction encoding.

The basic design requirements for a suitable homophonic encoder, i.e., the matrix $\mathbf{G}_H$, are pointed out above, and this section contains additional guidelines to design a dedicated homophonic encoding which provides maximum complexity of the underlying LPN problem for given implementation and communications overhead.

It is well known that the hardness of the LPN$_{n,\epsilon}$ problem in the average case, heavily depends on the parameter $\epsilon$ (see [5] and [7], for example). On the other hand, Theorem 2 implies that the parameter $\epsilon$ depends on the minimal value of the basic equations which should be linearly combined in order to eliminate the random variables from each equation of the system. Accordingly, Theorem 2 implies the following design criteria for construction of the matrix $\mathbf{G}_H$.

- **Weight.** For a given error-correcting code generator matrix $\mathbf{G}_{ECC}$, specify the homophonic code matrix $\mathbf{G}_H$ so that the resulting matrix $\mathbf{G}_H = [g_{ij}^{(H)}]_{i=1}^m{}_{j=1}^m$ satisfies:

$$\sum_{i=1}^{m-\ell} g_{\ell+i,m-\ell+j}^{(H)} \geq w \ , \ j = 1, m-\ell+2, \ldots, l \ ,$$

where $w$ is a parameter.

- **Dependability.** According to (14)-(16), the sub-matrix of the matrix $\mathbf{G}$ consisting of its $m - \ell$ last rows should be such that any of the columns is a linear combination of at least $w$ other columns. Consider thus the sub-matrix $\mathbf{G}^*$ determined by the rows $m - \ell + 1, m - \ell + 2, \ldots, m$ and columns $1, 2, \ldots, n$ of the matrix $\mathbf{G}$. We require that no column of the matrix $\mathbf{G}^*$ is equal to a linear combination of $w$ or less other columns of $\mathbf{G}^*$. This can be rephrased by asking

$$\operatorname{rank}(\mathbf{G}^*) \geq w + 1.$$

## 4.2 Implementation design criteria

On the sender side, both the homophonic and error-correcting encodings are performed via a single vector-matrix multiplication employing the matrix $\mathbf{G} = \mathbf{G}_H \mathbf{G}_{CEE}$.

On the receiver side the error-correction decoding and the homophonic decoding should be performed independently. First the errors should be corrected by the error-correction decoding, because the homophonic decoding requires error-free decoding input.

This implies that in order to minimize the implementation complexity, a desirable property is sparseness of the related matrices $\mathbf{G}_H$, $\mathbf{G}$ and $\mathbf{G}_H^{-1}$.

- **Sparsity.** For a given error-correcting code generator matrix $\mathbf{G}_{ECC}$, and a given security parameter $w$, specify the homophonic encoding matrix $\mathbf{G}_H$ in such a manner that either it is sparse or the resulting matrix $\mathbf{G}$ is sparse to provide minimization of the implementation complexity on the sender side, and at the same time the matrix $\mathbf{G}_H^{-1}$ is sparse in order to avoid too high computation overhead for the receiver.

# 5 Homophonic Code Constructions

Let us first write the $m \times m$ wire-tap matrix $\mathbf{G}_H$ and the $m \times n$ error-correcting matrix $\mathbf{G}_{ECC}$ as

$$\mathbf{G}_H = \left[\begin{array}{cc} \mathbf{G}_H^{(1)} & \mathbf{G}_H^{(2)} \\ \mathbf{I}_{m-l} & \mathbf{G}_H^{(4)} \end{array}\right], \ \mathbf{G}_{ECC} = \left[\begin{array}{c} \mathbf{G}_{ECC}^{(1)} \\ \mathbf{G}_{ECC}^{(2)} \end{array}\right] \tag{18}$$

where $\mathbf{G}_H^{(1)}$ is an $l \times (m-l)$ matrix, $\mathbf{G}_H^{(2)}$ is an $l \times l$ matrix, $\mathbf{I}_{m-l}$ denotes the $(m-l) \times (m-l)$ identity matrix, $\mathbf{G}_H^{(4)}$ is an $(m-l) \times l$ matrix written as

$$\mathbf{G}_H^{(4)} = \left[\begin{array}{cccc} g_{\ell+1,m-\ell+1}^{(H)} & g_{\ell+1,m-\ell+2}^{(H)} & \cdots & g_{\ell+1,m}^{(H)} \\ g_{\ell+2,m-\ell+1}^{(H)} & g_{\ell+2,m-\ell+2}^{(H)} & \cdots & g_{\ell+2,m}^{(H)} \\ \vdots & \vdots & \cdots & \vdots \\ g_{m,m-\ell+1}^{(H)} & g_{m,m-\ell+2}^{(H)} & \cdots & g_{m,m}^{(H)} \end{array}\right],$$

$\mathbf{G}_{ECC}^{(1)}$ is an $(m-l) \times n$ matrix and finally $\mathbf{G}_{ECC}^{(2)}$ is an $l \times n$ matrix, so that

$$\begin{aligned}\mathbf{G} &= \mathbf{G}_H \mathbf{G}_{ECC} \\ &= \begin{bmatrix} \mathbf{G}_H^{(1)} & \mathbf{G}_H^{(2)} \\ \mathbf{I}_{m-l} & \mathbf{G}_H^{(4)} \end{bmatrix} \begin{bmatrix} \mathbf{G}_{ECC}^{(1)} \\ \mathbf{G}_{ECC}^{(2)} \end{bmatrix} \\ &= \begin{bmatrix} \mathbf{G}_H^{(1)} \mathbf{G}_{ECC}^{(1)} + \mathbf{G}_H^{(2)} \mathbf{G}_{ECC}^{(2)} \\ \mathbf{G}_{ECC}^{(1)} + \mathbf{G}_H^{(4)} \mathbf{G}_{ECC}^{(2)} \end{bmatrix}.\end{aligned}$$

## 5.1 A generic construction

We now give a general construction method for the matrix $\mathbf{G}_H$. Choose first

$$\mathbf{G}_H^{(1)} = \mathbf{0}_{l \times (m-l)}, \ \mathbf{G}_H^{(2)} = \mathbf{I}_l.$$

Let us check that we already satisfy the information theoretical requirements.

- **Invertibility.** Since $\mathbf{G}_H$ is a square matrix, we can rephrase its invertibility using its determinant by asking
  $$\det(\mathbf{G}_H) \neq 0.$$
  Using Schur complement, this is equivalent to
  $$\det(\mathbf{G}_H^{(2)} - \mathbf{G}_H^{(1)} \mathbf{G}_H^{(4)}) \neq 0.$$
  The above choice of $\mathbf{G}_H^{(1)}$ and $\mathbf{G}_H^{(2)}$ gives $\det(\mathbf{I}_l) \neq 0$ which always holds, so that the invertibility condition is taken care of.

- **Security.** The matrix $\mathbf{G}_H$ should map $[\mathbf{a}||\mathbf{u}]$ so that in the resulting vector each bit of data from $\mathbf{a}$ is affected by at least one random bit from $\mathbf{u}$. Since
  $$[\mathbf{a}||\mathbf{u}] \begin{bmatrix} \mathbf{0}_{l \times (m-l)} & \mathbf{I}_l \\ \mathbf{I}_{m-l} & \mathbf{G}_H^{(4)} \end{bmatrix} = [\mathbf{u}, \mathbf{a} + \mathbf{u}\mathbf{G}_H^{(4)}],$$
  it is enough that $\mathbf{G}_H^{(4)}$ has no column with only zeroes to get that indeed each bit of data from $\mathbf{a}$ is affected by at least one random bit from $\mathbf{u}$.

We next look at the conditions coming from computational security. The weight condition can be rephrased as requiring that each column of $\mathbf{G}_H^{(4)}$ has Hamming weight at least $w$, which automatically makes sure that $\mathbf{G}_H^{(4)}$ has no column with only zeroes.

The dependability condition relates to the sub-matrix $\mathbf{G}^*$ determined by the rows $m-\ell+1, m-\ell+2, \ldots, m$ and columns $1, 2, \ldots, n$ of the matrix $\mathbf{G}$. Since $m-l$ counts the number of random bits, it is reasonable to assume that

$$m - l \leq l \iff m \leq 2l,$$

that is we use at most as many random bits as data bits. Since

$$\mathbf{G} = \mathbf{G}_H \mathbf{G}_{ECC} = \begin{bmatrix} \mathbf{0}_{l \times (m-l)} & \mathbf{I}_l \\ \mathbf{I}_{m-l} & \mathbf{G}_H^{(4)} \end{bmatrix} \begin{bmatrix} \mathbf{G}_{ECC}^{(1)} \\ \mathbf{G}_{ECC}^{(2)} \end{bmatrix} = \begin{bmatrix} \mathbf{G}_{ECC}^{(2)} \\ \mathbf{G}_{ECC}^{(1)} + \mathbf{G}_H^{(4)} \mathbf{G}_{ECC}^{(2)} \end{bmatrix}$$

with, assuming w.l.o.g that $\mathbf{G}_{ECC}$ is in systematic form,

$$\mathbf{G}_{ECC}^{(1)} = \begin{bmatrix} g_{1,1}^{(ECC)} & g_{1,2}^{(ECC)} & \cdots & g_{1,n}^{(ECC)} \\ g_{2,1}^{(ECC)} & g_{2,2}^{(ECC)} & \cdots & g_{2,n}^{(ECC)} \\ \vdots & \vdots & \cdots & \vdots \\ g_{m-\ell,1}^{(ECC)} & g_{m-\ell,2}^{(ECC)} & \cdots & g_{m-\ell,n}^{(ECC)} \end{bmatrix} = \begin{bmatrix} \mathbf{I}_{m-l} & \mathbf{0}_{(m-l) \times l} & \mathbf{P}_{(m-l) \times (n-m)} \end{bmatrix}$$

and
$$\mathbf{G}_{ECC}^{(2)} = \begin{bmatrix} g_{m-\ell+1,1}^{(ECC)} & g_{m-\ell+1,2}^{(ECC)} & \cdots & g_{m-\ell+1,n}^{(ECC)} \\ g_{m-\ell+2,1}^{(ECC)} & g_{m-\ell+2,2}^{(ECC)} & \cdots & g_{m-\ell+2,n}^{(ECC)} \\ \vdots & \vdots & \cdots & \vdots \\ g_{m,1}^{(ECC)} & g_{m,2}^{(ECC)} & \cdots & g_{m,n}^{(ECC)} \end{bmatrix} = \begin{bmatrix} \mathbf{0}_{l\times(m-l)} & \mathbf{I}_l & \mathbf{Q}_{l\times(n-m)} \end{bmatrix},$$

we can write the $l \times n$ matrix $\mathbf{G}^*$ as
$$\mathbf{G}^* = \begin{bmatrix} g_{2m-2l+1,1}^{(ECC)} & g_{2m-2l+1,2}^{(ECC)} & \cdots & g_{2m-2l+1,n}^{(ECC)} \\ \vdots & \vdots & \cdots & \vdots \\ g_{m,1}^{(ECC)} & g_{m,2}^{(ECC)} & \cdots & g_{m,n}^{(ECC)} \\ \mathbf{G}_{ECC}^{(1)} & +\mathbf{G}_H^{(4)}\mathbf{G}_{ECC}^{(2)} & & \end{bmatrix}$$
$$= \begin{bmatrix} \mathbf{0}_{(2l-m)\times(2m-3l)} & \mathbf{I}_{2l-m} & \mathbf{0}_{(2l-m)\times l} & \mathbf{P}'_{(2l-m)\times(n-m)} \\ \mathbf{G}_{ECC}^{(1)} + \mathbf{G}_H^{(4)}\mathbf{G}_{ECC}^{(2)} & & & \end{bmatrix}.$$

Now
$$\mathbf{G}_H^{(4)}\mathbf{G}_{ECC}^{(2)} = \mathbf{G}_H^{(4)}[\mathbf{0}_{l\times(m-l)}\ \mathbf{I}_l\ \mathbf{Q}_{l\times(n-m)}] = [\mathbf{0}_{m-l}\ \mathbf{G}_H^{(4)}\ \mathbf{G}_H^{(4)}\mathbf{Q}_{l\times(n-m)}]$$
so that finally
$$\mathbf{G}^* = \begin{bmatrix} \mathbf{0}_{(2l-m)\times(2m-3l)} & \mathbf{I}_{2l-m} & \mathbf{0}_{(2l-m)\times l} & \mathbf{P}'_{(2l-m)\times(n-m)} \\ \mathbf{I}_{m-l} & & \mathbf{G}_H^{(4)} & \mathbf{G}_H^{(4)}\mathbf{Q}_{l\times(n-m)} + \mathbf{P}_{(m-l)\times(n-m)} \end{bmatrix}.$$

The requirement is that
$$\text{rank}(\mathbf{G}^*) \geq w+1.$$
Since $l \leq m < n$, the rank of $\mathbf{G}^*$ is at most $l$, and it is enough to look at the rank of the $l \times m$ submatrix
$$\begin{bmatrix} \mathbf{0}_{(2l-m)\times(2m-3l)} & \mathbf{I}_{2l-m} & \mathbf{0}_{(2l-m)\times l} \\ \mathbf{I}_{m-l} & & \mathbf{G}_H^{(4)} \end{bmatrix} \tag{19}$$
which varies from $m-l$ to $l$ since the first $m-l$ columns are linearly independent. Thus if $w+1 \leq m-l$, the dependency condition is satisfied naturally. Otherwise, we need to build $\mathbf{G}_H^{(4)}$ such that $k$ of its columns, $k = 1, \ldots, 2l-m$ are linearly independent from the $m-l$ first columns of the above matrix. To do so, it is enough to consider the $2l-m$ first columns of $\mathbf{G}_H^{(4)}$, and we consider the truncated matrix (19)
$$\begin{bmatrix} \mathbf{0}_{(2l-m)\times(2m-3l)} & \mathbf{I}_{2l-m} & \mathbf{0}_{(2l-m)\times 2l-m} \\ \mathbf{I}_{m-l} & & \mathbf{A} \end{bmatrix}$$
where $\mathbf{A}$ contains the $2l-m$ first columns of $\mathbf{G}_H^{(4)}$. Let us further write
$$\mathbf{A} = \begin{bmatrix} \mathbf{A}_1 \\ \mathbf{A}_2 \end{bmatrix},$$
where $\mathbf{A}_1$ is a $(2m-3l) \times (2l-m)$ matrix, and $\mathbf{A}_2$ is a square $2l-m$ matrix. To control the rank of $\mathbf{G}^*$, we set $\mathbf{A}_1 = \mathbf{0}$ and we get
$$\text{rank}(\mathbf{G}^*) = m-l+k$$
where $k$ is the number of columns of $\mathbf{A}_2$ which are linearly independent from the the matrix
$$\begin{bmatrix} \mathbf{0}_{(2l-m)\times(2m-3l)} & \mathbf{I}_{2l-m} \\ \mathbf{I}_{m-l} & \end{bmatrix}. \tag{20}$$
Setting $\mathbf{A}_2$ to zero makes this computation easier. Indeed, to get $k$ such columns, it is enough to pick $k$ columns from the $2l-m$ identity matrix. This might give some columns with zero or very few ones, which looks like contradicting the weight condition. However, columns with higher Hamming weight can be easily obtained by taking linear combinations of the columns which will not change the rank.

- **Dependability and Weight.** Since

$$\text{rank}(\mathbf{G}^*) = m - l + k$$

where $k$ is the number of columns of the sub-matrix of $\mathbf{G}_H^{(4)}$ formed by taking its first $2m - l$ columns and last $2m - l$ rows which are linearly independent from (20), it is enough to ask for

$$k \geq w + 1 + l - m.$$

To ensure that each column of $\mathbf{G}_H^{(4)}$ has Hamming weight $w$, it is enough to take linear combinations of the columns.

- **Sparsity.** The choice of $\mathbf{G}_H^{(1)} = \mathbf{0}_{l \times (m-l)}$ and $\mathbf{G}_H^{(2)} = \mathbf{I}_l$ makes the $l$ first rows of $\mathbf{G}_H$ as sparse as possible, since removing any zero would make the matrix non-invertible anymore. Furthermore, the way the dependability condition is constructed is also optimal in the sense that it starts with the least number of 1 to get the wanted rank, and then obtains the desired Hamming weight of each column by linear combinations.

## 5.2 Examples of Constructions

Take $m = 2l$ so that $m - l = l$, and

$$\mathbf{G}_H = \begin{bmatrix} \mathbf{G}_H^{(1)} & \mathbf{G}_H^{(2)} \\ \mathbf{I}_l & \mathbf{G}_H^{(4)} \end{bmatrix} = \begin{bmatrix} \mathbf{0}_l & \mathbf{I}_l \\ \mathbf{I}_l & \mathbf{I}_l \end{bmatrix}.$$

The matrix $\mathbf{G}_H$ is clearly invertible. The Hamming weight of each column of $\mathbf{G}_H^{(4)}$ is 1, thus $w$ must be taken to be 0 or 1. The matrix $\mathbf{G}_H^{(2)}$ is chosen to be zero for increasing the sparsity of $\mathbf{G}_H$.

As a toy example, let us consider the $(7, 4)$ Hamming code with

$$\mathbf{G}_{ECC} = \begin{pmatrix} 1 & 0 & 0 & 0 & 1 & 1 & 0 \\ 0 & 1 & 0 & 0 & 1 & 0 & 1 \\ 0 & 0 & 1 & 0 & 0 & 1 & 1 \\ 0 & 0 & 0 & 1 & 1 & 1 & 1 \end{pmatrix},$$

and

$$\mathbf{G}_H = \begin{pmatrix} 0 & 0 & 1 & 0 \\ 0 & 0 & 0 & 1 \\ 1 & 0 & 1 & 0 \\ 0 & 1 & 0 & 1 \end{pmatrix}.$$

We have that

$$\mathbf{G}_H \mathbf{G}_{ECC} = \begin{pmatrix} 0 & 0 & 1 & 0 & 0 & 1 & 1 \\ 0 & 0 & 0 & 1 & 1 & 1 & 1 \\ 1 & 0 & 1 & 0 & 1 & 0 & 1 \\ 0 & 1 & 0 & 1 & 0 & 1 & 0 \end{pmatrix}.$$

The matrix $\mathbf{G}^*$ is thus

$$\mathbf{G}^* = \begin{pmatrix} 1 & 0 & 1 & 0 & 1 & 0 & 1 \\ 0 & 1 & 0 & 1 & 0 & 1 & 0 \end{pmatrix}.$$

Since the requirement is that the rank of $\mathbf{G}^*$ is at least $w$, this is clearly satisfied here since $w = 1$. Note that

$$\mathbf{G}_H^{-1} = \begin{pmatrix} 1 & 0 & 1 & 0 \\ 0 & 1 & 0 & 1 \\ 1 & 0 & 0 & 0 \\ 0 & 1 & 0 & 0 \end{pmatrix},$$

and the cost of encoding and decoding the homophonic code is the same.

In order to increase $w$, we could alternatively take

$$\mathbf{G}_H = \begin{pmatrix} 0 & 0 & 1 & 0 \\ 0 & 0 & 0 & 1 \\ 1 & 0 & 1 & 1 \\ 0 & 1 & 1 & 1 \end{pmatrix},$$

for which $w$ can be taken to be 2. Then, continuing with the $(7,4)$ Hamming code, we get

$$\mathbf{G}_H \mathbf{G}_{ECC} = \begin{pmatrix} 0 & 0 & 1 & 0 & 0 & 1 & 1 \\ 0 & 0 & 0 & 1 & 1 & 1 & 1 \\ 1 & 0 & 1 & 1 & 0 & 1 & 0 \\ 0 & 1 & 1 & 1 & 0 & 0 & 1 \end{pmatrix},$$

where

$$\mathbf{G}^* = \begin{pmatrix} 1 & 0 & 1 & 1 & 0 & 1 & 0 \\ 0 & 1 & 1 & 1 & 0 & 0 & 1 \end{pmatrix}$$

has rank $w = 2$ as required. This time

$$\mathbf{G}_H^{-1} = \begin{pmatrix} 1 & 1 & 1 & 0 \\ 1 & 1 & 0 & 1 \\ 1 & 0 & 0 & 0 \\ 0 & 1 & 0 & 0 \end{pmatrix},$$

and as expected, increasing $w$ correspondingly decreases the sparsity of $\mathbf{G}_H$ and its inverse.

## 6 Conclusion

The paper addresses the problem of design of a homophonic code for certain enhanced encoding-encryption based communication systems, reported and analyzed from information-theoretic point of view in [11]-[12]. The design is based on the guidelines implied by security evaluation from the computational complexity point of view. Accordingly, this paper yields the following: (i) security evaluation of the considered system from the computational complexity point of view; (ii) guidelines for design of a dedicated homophonic coding implied by the performed security evaluation; (iii) proposal of a dedicated homophonic code which provides the desired level of security and at the same time provides low implementation overhead.

The security evaluation of the employed encryption is considered by hardness of recovering the secret key based on the algebraic representation of the encryption in CPA scenario. It is shown that the addressed secret key recovery is at least as hard as the LPN problem when, assuming an appropriate design, the corrupting noise is $\epsilon = \frac{1-(1-2p)^{w+1}}{2}$ and $p < 0.5$ and $w$ are the system parameters. Note that the in the average complexity consideration, the LPN problem corresponding to the parameter $\epsilon$ is much harder than the one with the parameter $p$. Accordingly, assuming that the parameters of the scheme are appropriately selected, the complexity of the secret key recovery based on the algebraic representation appears approximately as hard as the exhaustive search over all possible secret keys.

The results of security evaluation are considered as guidelines for design of a dedicated homophonic encoder which provides a desired security level and minimize the implementation complexity. Assuming that the homophonic code should be a linear one, beside the basic requirement on the invertibility and the mixing properties of the generator matrix, the following three additional criteria are pointed out and specified in Sections 4.1 and 4.2: (a) Weight on columns of the generator matrix; (b) rank of the generator matrix; (c) sparsity of the generator matrix. The criteria (a) and (b) appear as an implication of the security requirements, and the criterion (c) is related to minimization of the implementation overhead. The previous design criteria are employed for design of a dedicated homophonic code. A generic design of the homophonic coding dedicated to the considered security enhanced communication system is proposed and it is shown that the design fulfills all the given criteria.

## Acknowledgments

The research of F. Oggier is supported in part by the Singapore National Research Foundation under Research Grant NRF-RF2009-07 and NRF-CRP2-2007-03, and in part by the Nanyang Technological University under Research Grant M58110049 and M58110070. This work was done partly while M. Mihaljević was visiting the division of mathematical sciences, Nanyang Technological University, Singapore, and partly while F. Oggier was visiting the Research Center for Information Security, Tokyo. M. Mihaljević is partly supported via the Project # 174008.